\documentclass[pre,preprintnumbers,twocolumn]{revtex4}
\usepackage{graphicx,t1enc}
\usepackage{epsfig}
\usepackage{amssymb,amsmath,amsthm}

\begin{document}

\title{The effect of the topology on the spatial ultimatum game}

\author{M. N. Kuperman}
\affiliation{Centro At{\'o}mico Bariloche and Instituto Balseiro,
8400 S. C. de Bariloche, Argentina \\
Consejo Nacional de Investigaciones Cient{\'\i}ficas y
T{\'e}cnicas, Argentina}

\author{S. Risau Gusman}
\affiliation{Centro At{\'o}mico Bariloche and Instituto Balseiro,
8400 S. C. de Bariloche, Argentina \\
Consejo Nacional de Investigaciones Cient{\'\i}ficas y
T{\'e}cnicas, Argentina}

\begin{abstract}

In this work we present an analysis of a spatially non homogeneous
ultimatum game. By considering different underlying topologies as
substrates on top of which the game takes place we obtain
nontrivial behaviors for the evolution of the strategies of the
players. We analyze separately the effect of the size of the
neighborhood and the spatial structure. Whereas this last effect
is the most significant one, we show that even for disordered
networks and provided the neighborhood of each site is small, the
results can be significantly different from those obtained in the
case of fully connected networks.
\end{abstract}

\maketitle

\section{Introduction}

In the last years, Game Theory has been recognized as a powerful
alternative way to examining economics \cite{cam1,gtb}. The models
analyzed under this scheme consists of sets of agents that play a
certain game and a set of the strategies that can be used by the
agents. Game theory can be understood as a mathematical tool for
analyzing and predicting human behavior in strategic situations.
In the last years many physicist have directed their attention
towards the analysis of several market games
\cite{ber,tan,hua,gua,per1,jef}

The equilibrium analysis assume that all the players display
strategic thinking and optimizing behavior. However, it is widely
accepted, and has been experimentally shown, that not every player
behaves in a rational way. The realization of this difference led
to the creation of Behavioral Economics, a branch of economics
closely related to the study of the behavior of economical agents
rather than economical quantities \cite{beco1,cam2}. One of the
most interesting results obtained in this field is the observation
that real individuals do not behave according to the classical
assumptions of {\it homo economicus} \cite{per}, a completely rational
individual who seeks to optimize his utilities with the least
possible cost.

The gap between economics and Game Theory has been bridged by
Evolutionary Game Theory, which takes into account the possibility
that the strategies of the agents can change following some
evolutionary rule. The Evolutionary Game Theory has succeeded in
explaining how cooperation can arise in populations playing
non-cooperative games, i.e. in games where cooperation is possible
but is not favored. Included in the group of non-cooperative games
with economical interest we find the Prisoner's Dilemma, the
Ultimatum Game, bargaining games, etc.

In the present work we focus on some aspects related to the
Ultimatum Game. The essential features of this game can be very
easily summarized. Two individuals are told that they have the
opportunity to split a given amount of externally provided money.
One of the individuals is randomly chosen as the first player and to
assume the role of the offerer. He/she has to make a one time
offer, consisting in indicating how much of the total amount of
money is to be given to each player. The other player, as the
respondent, has the opportunity to either accept or reject this
offer. If the offer is rejected both get nothing. If the offer is
accepted, each one gets the accorded part. Both participants are
aware of the rules of the game before making any decision.

The Ultimatum Game is a particular case of bargaining. Game theory
predicts that the best strategy is to offer an unequal split
favoring the offerer. In \cite{rub} it is shown that if $\epsilon$
is the lowest allowed partition and given that a rational
responder will prefer a small amount to nothing, the best strategy
for the offerer is to give just $\epsilon$ and take the rest. But
studies made by behavioral economists have shown that most of the
time real individuals tend to reject unequal offers. The first
studies are described in \cite{gss}. Since then there have been
extensive studies on the behaviors of the players under different
circumstances and within a wide spectra of cultural environments.
Their results do not lead to a unique behavioral profile, and in
particular they show clearly that human players usually do not act
as the {\it homo economicus}
\cite{gss,kkt1,kkt2,hen2,and,hen1,sol,bre}.

It has also been shown (\cite{bros}) that inequity aversion may
not be a exclusively human feature: brown capuchin monkeys ({\it Cebus
Capella}) seem to respond negatively to unequal reward distribution
in exchanges with a human experimenter.

In order to explore the ultimatum game beyond the "static"
formulation by Rubinstein \cite{rub}, some authors have analyzed
the evolutionary ultimatum game \cite{now,now2,now3,kil,anxo}. In
these works it is shown that when the agents are placed in an
ordered network (and therefore constrained to play with, and
imitate, only their neighbors) the evolution takes the system to
more equitable strategies than predicted by the rational players
hypothesis. One natural question that arises is whether this
effect is only due to the spatial distribution of the players or
it is also due to the fact that the players are restricted to play
and imitate only a very small number of agents.

One of the goals of this work is to stress the important role
played by the underlying topology. Notice that ordered networks
differ from the fully mixed case not only in that agents are
connected to a very small set of other agents, but also on the
fact that the clustering is much smaller, i.e. the neighbors of a site
are not necessarily connected among themselves. In this work we
analyze these features separately to understand the effect they
have on the evolution of strategies.

We show that some field results can be in correspondence with our
findings. For example, in \cite{hen2} it is shown that the
behavior of players cannot be univocally associated neither with
the rational nor the altruistic attitudes. On the contrary,
experiments across different cultural environments show that it is
spread over a wide spectra of possibilities. This result was
mentioned but not discussed in previous works. At the same time we
establish interesting relationships between the outcome of an
evolutionary situation and the underlying social topology.

\section{The model}

The model consists of a set of $N$ players located on a network,
which defines the neighborhood of each player, i.e. the subset of
the whole population that is available for interaction. Each
player $i$ is assigned a playing strategy that consists of a pair
of real numbers $(o_i,a_i)$ within the interval $[0,1]$. An
interaction consists in taking a pair of linked players, and let
them play twice, alternating the roles of offerer and respondent.
$o_i$ is the offer of player $i$ when acting as offerer, and $a_i$
is the smallest amount $i$ accepts when acting as respondent. The
total sum allotted in each game is $1$.

The temporal evolution of the game is organized into generations.
In each generation, each player interacts with all of its
neighbors. The payoff of $i$ when playing with $j$, $w_{ij}$ is

\begin{equation} w_{ij}= \left\{ \begin{array}{ll}
1-o_i+o_j & \mbox{\ \ \  if \ } o_i
\ge a_i \mbox{\ and\ } o_j\ge a_i \\
1-o_i & \mbox{\ \ \  if \ } o_i
\ge a_i \mbox{\ and\ } o_j< a_i \\
o_j&  \mbox{\ \ \  if \ } o_i
< a_i \mbox{\ and\ } o_j\ge a_i \\
0 &  \mbox{\ \ \  if \ } o_i
< a_i \mbox{\ and\ } o_j<  a_i \\
\end{array}
\right. \label{pyf}
\end{equation}

After each agent has played with its entire neighborhood we
compute the cumulative payoff of each individual and consider that
a game generation has concluded. It is at this moment that the
evolutionary dynamics takes place. In the next generation all the
players are replaced by their offspring. The strategy of a site is
updated by choosing one strategy in the neighborhood (including
the site to be updated) with probability equal to the ratio
between the individual cumulative payoff and the total cumulative
payoff of all the sites in the neighborhood.

This warrants a competition process in which only the fittest or
more successful strategies survive. In the next generation, a new
breed of players occupies the sites of the network, with reset
payoffs but with strategies determined by the outcome of the
previous generation. The fact that a strategy was successful in a
given generation does not guarantee its success in the next one,
with a different distribution of strategies.

To avoid the system to get stuck in spurious local minima, we add
some noise in the form of small mutations, associated to a
mutation rate $\epsilon$ \cite{now2}. Once  the process of reproduction is
finished by determining which player will leave its offspring in
which sites, the descendants copy their ancestor strategy with a
small variation: if the individual that formerly occupied the site
$m$ leaves a descendant in the place $l$, the strategy of the new
occupant of the site $l$ is then
\begin{equation}
(o_l(t+1),a_l(t+1))=(o_m(t)+\delta_o,a_m(t)+\delta_a)\end{equation}
with $\delta \in [-\epsilon,\epsilon]$ a real random number.

We have performed simulations in three different topologies:
ordered, disordered, and $k$-Small World Networks ($k$-SWN), to
interpolate smoothly between the ordered and disordered
topologies. The ordered topology consists of nodes on a ring,
joined to their first $k$ neighbors to each side. These networks
are highly clustered: many neighbors of each node are connected
among themselves, forming triangles. This characteristic is
quantified by the clustering coefficient which is the number of
triangles centered on each node divided by the number of pairs of
neighbors, averaged over all the nodes. The disordered topology we
consider is a random graph where all the nodes have the same
degree $2k$ (also called regular random graphs). The third topology
is a variation of the small world networks of Watts and Strogatz
(WS)\cite{ws}.

The algorithm of construction of WS networks is as follows:
starting from an ordered network, the ring is traversed and with
probability $p'$ each link is rewired to a random node. Double and
self links are not allowed. Though the algorithm conserves the
total number of links, at the end of the process the degree of
each node is statistically characterized by a binomial
distribution, for $p'>1$. As we are interested in filtering any
effect related to changes in the size of the neighborhoods we
modify the original WS algorithm to constrain the resulting
networks to a subfamily with a delta shaped degree distribution.
We call this family of networks the $k$-Small World Networks
($k$-SWN), where $2k$ indicates the degree of the nodes. We start
again from an ordered network and define a disorder parameter $p$
that plays a role analogous to that of $p'$ in WS networks. To
proceed with the reconnection of the network we choose two couples
of linked nodes (or partners) rather than one. With probability
$p$ we decide whether to switch or not the partners in order to
get two new pairs of coupled links. In this way all the nodes
preserve their degree while the process of reconnection assures
the introduction of a certain degree of disorder. It must be
stressed that the dependence with $p$ of the clustering
coefficient and path length is qualitatively similar to what is
observed as a function of $p'$ in SWNs.

\begin{figure}[!hbt] \centering \resizebox{9cm}{!}
{
\includegraphics{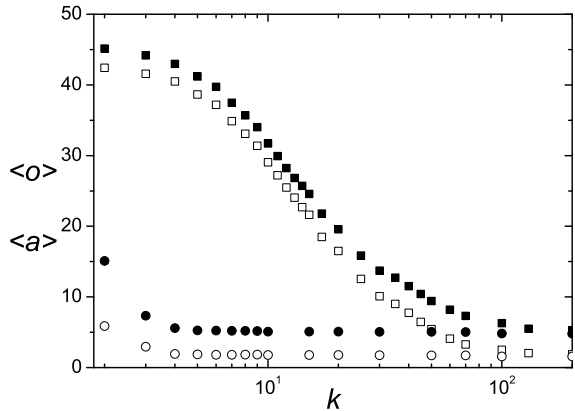}
}
\caption{Asymptotic mean thresholds $<o>$ (full symbols) and $<a>$
(empty symbols) for agents in a k-small world network, as a
function of the node degree $k$.  Squares: Ordered Network,
Circles: Regular Random Network} \label{vsk}
\end{figure}

\begin{figure}[!hbt] \centering \resizebox{9cm}{!}
{\rotatebox[origin=c]{0}{
\includegraphics{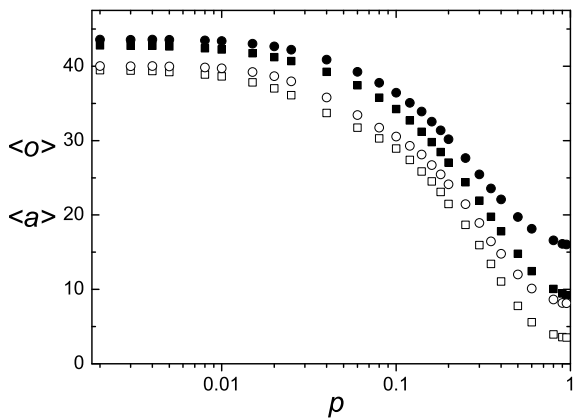}}}
\caption{Asymptotic mean thresholds $<o>$ (full symbols) and $<a>$
(empty symbols) for agents in a k-small world network, as a
function of the disorder parameter of the network. Squares: $k=2$,
Circles: $k=3$} \label{vsp}
\end{figure}

\section{Numerical results}

The simulations were done on networks with $10^3$ - $10^4$
individuals and different values of $k$. Defining a  generation as
the situation when each player plays twice (as offerer and as respondent) with all his/her neighbors
we performed time averages over the last $10^6$ generations of each realization, after a transient equally long.
At the same time, the time averaged results for a given set of parameters
$p$ and $k$ were obtained after averaging over 100 individual cases.

We begin by showing the comparison (see Fig 1) between the
strategies attained by agents placed in a ring, each one connected
to its first $k$ neighbors, and the ones attained when playing in
a {\it regular random graph} (i.e. a graph where each node is
connected to exactly $k$ random neighbors). These last networks
have a vanishing clustering coefficient, which means that,
locally, they are isomorphic to trees. On the other hand, the
networks defined on the ring have a clustering that increases with
$k$: $c=\frac{3}{4} \frac{k-2}{k-1}$. It is interesting to see
that evolution in a completely disordered network can also lead to
strategies that are rather far from the rational prediction,
provided that the number of neighbors is small enough. On the
other hand, a small increase in the number of neighbors leads to
strategies much closer to the rational prediction. This effect is
much less pronounced in the case of the ring networks but it must
be borne in mind that clustering is also increasing in this case.
The effect of clustering can be grasped by comparing the curves
for fixed values of $k$: larger clustering seems to lead to the
evolution of more equitable strategies. To analyze this in more
detail we have performed simulations of the Ultimatum Game in
k-Small World Networks.

\begin{figure}[!h] \centering \resizebox{9cm}{!}
{\rotatebox[origin=c]{0}{
\includegraphics{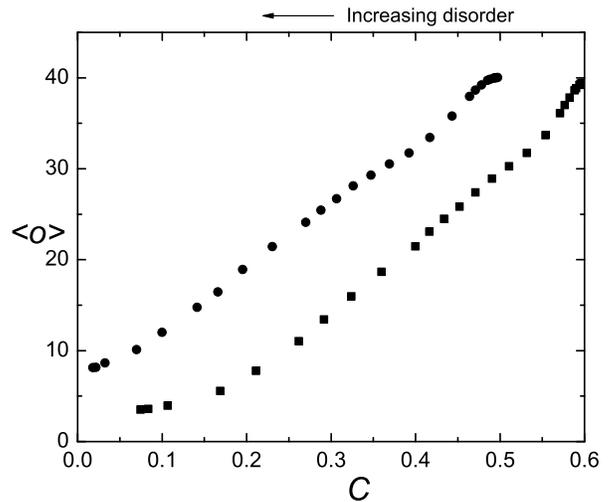}}}
\caption{Asymptotic mean thresholds $<o>$  for agents in a k-small
world network, as a function of the clustering of the network.
Squares: $k=2$, Circles: $k=3$} \label{ksw}
\end{figure}

As we vary the disorder parameter $p$ we observe a monotonous
smooth behavior of the values of $<o>$ and $<a>$. In Fig.
\ref{vsp} we show the mean values of $<o>$ as a function of the
disorder parameter $p$ ($<a>$ is not shown, as its behavior is
very similar). The behavior of this quantity looks qualitatively
very similar to that of the clustering coefficient. To stress this
fact we have plotted $<o>$ vs. $C$ in Fig. \ref{ksw}, where an
almost linear dependence can be observed.

\section {Invasion analysis}

When studying the dynamics of systems embedded in ordered or
semiordered networks it is difficult to go beyond numerical
simulations. Direct analytical approaches usually become
impossibly complicated in the presence of short loops. For this
reasons people have used indirect approaches with the hope of
capturing the essence of the problem.

One possibility to study the asymptotic state is to calculate the
strategy that a cluster of mutants should have to invade an
homogeneous population. The behavior of this as the topology is
changed has been shown to give some clues about the asymptotic
state of different systems \cite{Killingback99,Hauert01}. We have
performed these calculations for the limit cases of a completely
ordered and a completely disordered network. The calculation is
analogous to the one in \cite{now}, where only the conditions for
invasion of clusters of two and three sites were considered, for
the ordered network with $k=2$. In an infinite population of
individuals with offer and acceptance thresholds set to $o_b$ and
$a_b$, a cluster of $n$ mutants is introduced, with offer and
acceptance thresholds set to $o_m$ and $a_m$ with $o_m > a_m > o_b
> a_b$. For this initial state we derive the condition $o<o_c(k)$
that must be satisfied for the mutants to leave more than $n$
descendants in the next generation (this is only a first
approximation, because mutant expansion might not continue
indefinitely \cite{Hauert01}). Notice that for a homogeneous
population, with mutation allowed, the average thresholds should
be bigger that $o_c$. Otherwise a mutation could generate a mutant
cluster that would invade the population, thus raising the average
threshold.

\begin{figure}[!h] \centering \resizebox{9cm}{!}
{\rotatebox[origin=c]{0}{
\includegraphics{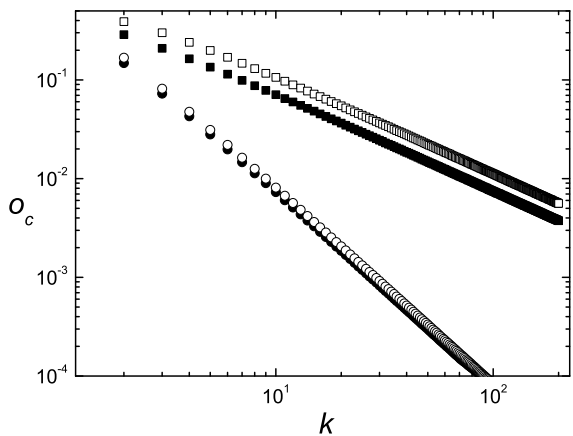}}}
\caption{Critical invasion thresholds as a function of the degree
of the nodes and for a cluster size 2 (full symbols) and 3 (empty
symbols). Squares: Ordered Network, Circles: Regular Random
Network} \label{inv}
\end{figure}

The behavior of $o_c(k)$ for clusters of two and three sites, in ordered and
completely disordered networks is shown in Fig \ref{inv}. The
decrease of both curves with $k$ was to be expected, because if a
cluster has a strategy well suited for invading the population,
increasing the number of its neighbors can only add to its
success. A more interesting feature is the relationship between
both curves for the same value of $k$. In this case the clusters
have the same number of connections, but they differ in how these
connections are arranged: in the completely disordered network all
these connections go to different agents, whereas in the ordered
network the two members of it are connected mostly to the same
sites (i.e. there are more triangles). Thus, the figure shows that
acting cooperatively leads to better reproductive chances, and this
is favored in networks with larger clustering coefficient. Notice
that the average payoff of the agents is the same in the two
cases. It is their relative fitness what changes.

The same features are present when one compares the invasion
thresholds for larger compact clusters, if their size does not
depend on $k$.

\section{Conclusions}

The rationale of the present work was to reveal the influence of
the underlying topology on the outcome of an evolutionary
Ultimatum Game. Though previous works have already addressed this
aspect, some interesting details could only be unveiled through a
more detailed analysis. This task was achieved by making numerical
simulations on several types of networks, with different amount of
disorder and degree distribution. We have observed that both the
increase of the neighborhood size and the increase of the degree
of disorder have a similar effect, leading a population of players
towards responses with increasing levels of "rationality". As the
increment of the size of the neighborhood makes a given network
converge towards a fully connected graph and thus being associated
to a mean field situation, the effect of the increase on the
disorder was not clear up to now. We have shown the transition in
the behavior of the population when the underlying topology varies
continuously between the extremes already studied in the
literature.

\begin{figure}[!h] \centering \resizebox{9cm}{!}
{\rotatebox[origin=c]{0}{
\includegraphics{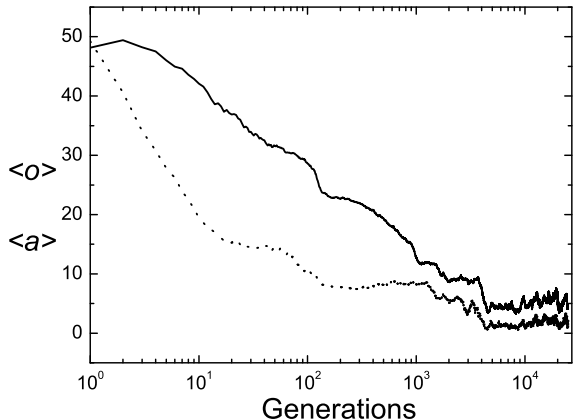}}}
\caption{Temporal behavior of the thresholds $<o>$ (full) and $<a>$
(dotted) in the fully connected network} \label{tmp}
\end{figure}

The behavior of the asymptotic state of the system as the
clustering of the underlying network changes can be compared to
that of some games where the evolution of cooperation is analyzed.
Individuals with fair strategies can survive provided that they
are surrounded by neighbors with similar behavior. A highly
clustered network favors this necessary condition and the system
naturally evolves to a situation when the mean level of offers is
around 50\%. On the contrary, when the underlying network present
a low level of clusterization, the mean values go to values close
to zero. In this case, the transient shows first a diminishing
acceptance threshold, that finally drives the offer value to lower
values as well, as shown in Figure \ref{tmp}. If we identify the
fair strategies as being more cooperative that those with lower
offers the results can be interpreted within the frame of the
phenomenology of other cooperative games where the spatial
distribution plays the same role as here and leads to the same
qualitative results.

So far, we have made a description and analysis of the obtained
results. After analyzing our results we are inclined to propose an
explanation for the emergence of a collective behavior that sets a
discrepancy between the theoretical predictions and field results
in relation to the ultimatum game. Furthermore, the experimental
results show a wide spectra of behaviors that could not be easily
explained or unified within a suitable frame.  In this work we
study an evolutionary process that changes the strategies of the
populations and can lead to the survival of the fittest behaviors.
The final composition of the evolutionary strategies of the
population seems to depend on the social underlying architecture.
This could explain, in principle, the fact that people with
different social organization display different collective
behaviors. The model succeeds to qualitatively reproduce the
complexity of the observed experimental results based on very
simple assumptions and interaction rules. Summarizing we show that
the whole spectra of observed behaviors can be reproduced by
considering an evolutionary process that together with the
subtleties of the social organization shapes the behavior of
different people.

\section{Appendix}

We consider a population of agents with offer and acceptance
thresholds set to $p_0$ and $q_0<p_0$, with two connected mutants
with thresholds $p_m > q_m > p_0 > q_0$. Agents can be connected
to none, one, or both mutants, depending on the topology of the
network. Their respective average payoff after a round of games
will be:

\begin{eqnarray}
f_0(k) &=& 2k \nonumber \\
f_1(k) &=& 2k-1+p  \nonumber \\
f_2(k) &=& 2k-2+2p
\end{eqnarray}

\noindent where $2k$ is the number of neighbors. For the mutants,
the payoff is:

\begin{equation}
f_m(k) = 1+(2k-1)(1-p)
\end{equation}

At the next generation, the new strategy for each site is chosen
randomly among its own and its neighbors strategies, with a
probability proportional to their respective average payoffs.
Thus, the expected number of mutants at the next generation is:

\begin{eqnarray}
m(k) &=& \frac{4f_m(k)}{2f_m(k)+(2k-1)f_1(k)}+ \nonumber \\ && \frac{(2k-1)f_m(k)}{f_m(k)+f_1(k)+(2k-1)f_0(k)} \nonumber \\
\end{eqnarray}

\noindent for the disordered case and

\begin{eqnarray}
m(k) &=& 2 \frac{2f_m(k)}{2f_m(k)+2(k-1)f_2(k)+f_1(k)} + \nonumber
\\ &&  \sum_{j=1}^{k-1}
\frac{2f_m(k)}{2f_m(k)+(2k-2-j)f_2(k)+f_1(k)+jf_0(k)} + \nonumber
\\  && \frac{f_m(k)}{f_m(k)+(k-1)f_2(k)+f_1(k)+kf_0(k)}
\end{eqnarray}

\noindent for the completely ordered case. The invasion condition
for $p$ can then be extracted from the equation $m(k)>2$.


\end{document}